\documentclass[a4paper,twoside,12pt]{article}

\usepackage{psfig}
\usepackage{amsfonts}
\usepackage{amssymb}

\newcommand{\ZZ}{\mathbb{Z}}

\title{Simple models with Alice fluxes} \author{J.Striet$^1$ and
F.A.Bais$^2$\\[2mm] Institute for Theoretical Physics \\ University of
Amsterdam \\ Valckenierstraat 65 \\ 1018XE Amsterdam \\ The
Netherlands\\}

\begin{document}
\maketitle
\date

\begin{abstract}
\noindent We introduce two simple models which feature an Alice
electrodynamics phase. In a well defined sense the Alice flux
solutions we obtain in these models obey first order equations similar
to those of the Nielsen-Olesen fluxtube \cite{NO} in the abelian higgs
model in the Bogomol'nyi limit. Some numerical solutions are presented
as well.\\
\end{abstract}
\footnotetext[1]{jelpers@science.uva.nl}
\footnotetext[2]{bais@science.uva.nl}
\section{Alice electrodynamics}
Alice Electrodynamics is a theory with gauge group $
H=U(1)\ltimes\ZZ_2$, i.e. an absolute minimally non-abelian extension
of ordinary electrodynamics, where charge conjugation has been turned
into a local symmetry. As this non-abelian extension is discrete, it
only effects 'electrodynamics' through certain global (topological)
features, such as the existence of Alice fluxes and Cheshire charges
\cite{schwarz,Alford}. In an Alice phase of some theory one has the
possibility of a topological stable flux tube or string coexisting
with an unbroken $U(1)$, because the connectivity of $H$ is
nontrivial: $\Pi_0(H) = \ZZ_2$. The Alice phase is usually obtained by
spontaneous breaking of a larger, continuous non-abelian symmetry
group. The original Alice model studied by Schwarz \cite{schwarz}, is
a $SU(2)$ theory spontaneously broken down to a $U(1)\ltimes\ZZ_2$ by
a higgs field in the $5-$dimensional representation of (see
also\cite{shankar}).  The higgs field is chosen in this
representation, because it is the smallest irreducible representation
which admits H as a residual symmetry group and allows for a single
valued vacuum configuration that supports Alice fluxes.

In this paper we will discuss two alternative models, which support an
Alice phase. Before doing so, we briefly review the salient features
of the model discussed in \cite{schwarz,shankar}.\\ The action is
given by:
\begin{equation}
S = \int d^4 x\ \{ Tr\ \{ -\frac{1}{8}F^{\mu\nu}F_{\mu\nu} -
\frac{1}{4}D^\mu \Phi D_\mu \Phi \ \} + V(\Phi)\}\;\;\; ,
\label{action}
\end{equation}
where the higgs field $\Phi = \Phi^{ab}$ is a real symmetric traceless
$3\times3$ matrix.\\ The most general renormalisable potential is given by:
\begin{equation}
V=-\frac{1}{2}\mu^2Tr\Phi^2 -\frac{1}{3}\gamma Tr\Phi^3 + \frac{1}{4}
\lambda (Tr\Phi^2)^2
\label{potential}
\end{equation}
By a suitable choice of parameters the higgs field will acquire a
vacuum expectation value, $\Phi_0$. In a gauge where $\Phi_0$ is
diagonal it takes the form $\Phi_0= diag(a,b,-a-b)$.
For a certain range of potential parameters one furthermore 
has that $a=b$, so that $\Phi_0$ is given by:
\begin{equation}
\Phi_0=\left( \begin{array}{ccc}
a&0&0\\
0&a&0\\
0&0&-2a
\end{array}\right)
\label{phi0}
\end{equation}
with $a=b=\frac{\gamma\pm\sqrt{\gamma^2+24\mu^2\lambda}}{12\lambda}$.
Indeed, this ground state is invariant under rotations around the
$T_3$-axis ($U(1)$) and invariant under rotations by an angle $\pi$
around any axis perpendicular to the $T_3$-direction ($\ZZ_2$). These
two transformations do not commute with each other, in fact they
anti-commute, so the resulting residual gauge group is indeed
$U(1)\ltimes\ZZ_2$. This means that we have Alice electrodynamics as
the low energy effective theory in this model.\\ An alternative way to
see the structure of the residual gauge group, is to think of the
higgs field as the symmetric traceless product of two vectors,
$\phi_1$ and $\phi_2$,
\begin{equation}
\Phi^{ab}=\phi_1^a\phi_2^b +
\phi_2^a\phi_1^b-\frac{2}{3}\delta^{ab}(\vec{\phi_1}\cdot\vec{\phi_2}).
\end{equation}
If both isovectors, $\vec{\phi_i}$, are non zero, there is in general
only a $\ZZ_2$ gauge symmetry left, $\vec{\phi_i}\to \vec{\phi_i'}
=-\vec{\phi_i}$. However, in case that both isovectors are
(anti-)parallel, the gauge group is $U(1)\ltimes\ZZ_2$. If one of the
isovectors is zero, the gauge group is not broken at all and the
symmetry remains $SU(2)$. These are the residual gauge groups which
one may encounter in this model. It is easy to
show that the case where the two isovectors are (anti-)parallel,
corresponds to the situation where  $\Phi=\Phi_0$ .

\subsection{The Alice flux solution}
\begin{figure}[t,h,b]
\psfig{figure=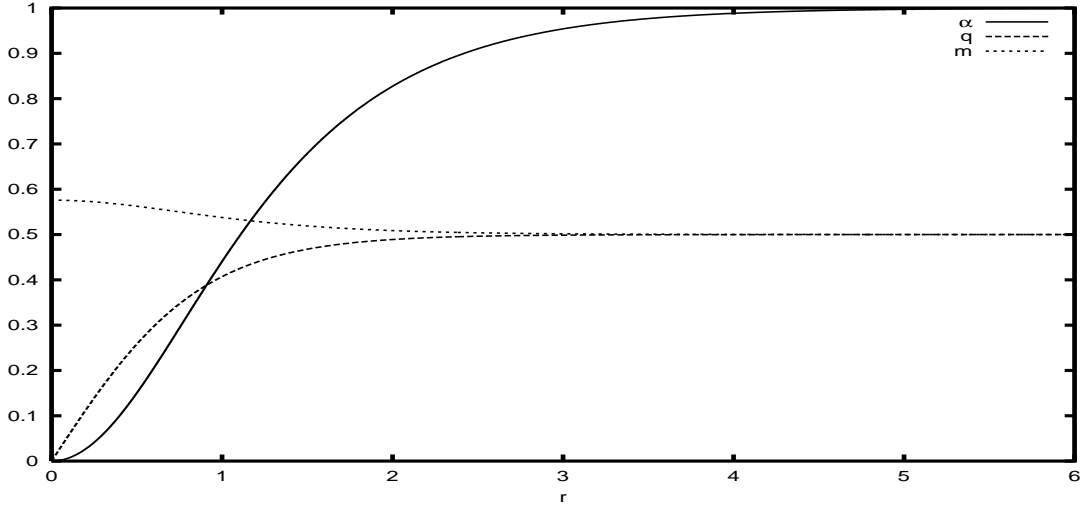,angle=-90,width=15cm,height=7cm}
\caption[somethingelse]
{\footnotesize A regular solution for the fields with an Alice flux for $\xi=1$ and $\chi=-1$.}
\label{shankar.eps}
\end{figure}

In this section we will present explicit regular solutions,
corresponding to an Alice flux tube along the z-axis, which where
constructed in \cite{postma}.\\ To have a static finite energy
solution, all terms in the energy density should go to zero at spatial
infinity. Thus the covariant derivatives need to vanish at spatial
infinity. Let's look at the angular derivative, the condition
$D_\theta \Phi=0$ tells us that the higgs field has the
following form at spatial infinity.
\begin{equation}
\Phi(\theta) = S(\theta) \Phi(0) S^{-1}(\theta)
\end{equation}
with
\begin{equation} 
S(\theta) = \exp\{e\int^\theta_0 rA_\theta d\theta \}
\label{Stheta}\end{equation}
Since we are looking for solutions which correspond to an Alice flux,
$S(2\pi)$ needs to be an element of the disconnected part of the
(residual) gauge group. A simple choice for $A_\theta$ doing this is
$A_\theta=\frac{1}{2er}T_1$.\\ This leads to the ansatz:
\begin{eqnarray}
A_\theta &=& \frac{\alpha(r)}{2er}T_1\\
\Phi(r,\theta)&=& e^{\frac{\theta T_1}{2}}\Phi(r)e^{-\frac{\theta T_1}{2}}
\end{eqnarray}
where the tensor $\Phi(r)$ is conveniently parameterized as,
\begin{equation}
\Phi(r)=m(r)\left(\begin{array}{ccc}
1&0&0\\
0&-\frac{1}{2}&0\\
0&0&-\frac{1}{2} 
\end{array} \right)
+
q(r)\left(\begin{array}{ccc}
0&0&0\\0&\frac{3}{2}&0\\
0&0&-\frac{3}{2} 
\end{array} \right)
\end{equation}
The part proportional to $m(r)$ is the part of the higgs field that is
invariant under rotations generated by $T_1$. The boundary condition
at spatial infinity is $m(\infty)=q(\infty)$, implying that
$\Phi(\infty)$ is of the form (\ref{phi0}), i.e. the residual symmetry
is $U(1)\ltimes\ZZ_2$ indeed, where the electrodynamic $U(1)$ is
generated by $T_3$. At the origin, $m$ and $q$ have to satisfy
different boundary conditions; the field $q(r)$ needs to go to
zero. The term proportional to $m(r)$ is invariant under $T_1$
rotations, therefore $m(r)$ does not need to go to zero. Again, this
means that the higgs field is of the form (\ref{phi0}), i.e. the
unbroken gauge group is $U(1)\ltimes\ZZ_2$. However, the unbroken
$U(1)$ is generated by $T_1$. Finally, the field $\alpha(r)$ needs to
be zero at the origin and unity at spatial infinity.\\ Inserting this
ansatz in the field equations gives, after suitable rescalings, the
following set of equations.
\begin{eqnarray}
\partial_r^2 \alpha(r) -\frac{1}{r}\partial_r\alpha(r) &=& 9
q^2(r)(\alpha(r)-1) \label{eom21}\\ \partial_r^2 q(r) +
\frac{1}{r}\partial_r q(r) &=& \frac{(\alpha(r)-1)^2q(r)}{r^2} +
\xi(9q^2(r)+3m^2(r)-2)q(r)\nonumber \\ &&+2\chi m(r)q(r)
\label{eom22}\\ \partial_r^2 m(r) + \frac{1}{r}\partial_r m(r) &=&
\xi(9q^2(r)+3m^2(r)-2)m(r)+\chi (3q^2(r)-m^2(r))
\label{eom23}
\end{eqnarray}
We summarize the boundary values for the rescaled fields below:\\[2mm]
\begin{displaymath}
\begin{tabular}{ccc}
\hline
field&$r\to0$ & r$ \to \infty$\\
\hline
$\alpha(r)$ & $0$ & $1$\\
$q(r)$ &$ 0$ &$q(\infty)$\\
$m(r)$ &$constant$  & $m(\infty)$\\
\hline
\end{tabular}
\end{displaymath}\\[2mm]
where,
\begin{equation}
m(\infty)=q(\infty)=\frac{-\chi\pm\sqrt{\chi^2+24\xi^2}}{12\xi}\equiv
a(\xi,\chi)
\label{a}
\end{equation}
with $\xi=\frac{\lambda}{e^2}$ and
$\chi=\frac{\gamma\sqrt{\lambda}}{\mu e^2}$.\\ The system (\ref{eom21}
- \ref{eom23}) was solved numerically with the help of a relaxation
method in \cite{postma}. The solution for the potential parameter
values $\xi=1$ and $\chi=-1$ is given in figure \ref{shankar.eps} .\\
The situation at hand is reminiscent to the one considered by Witten
\cite{witten} for a $U(1)\times\tilde{U}(1)$ model, in the sense that
we have an unbroken $U(1)\ltimes\ZZ_2$ at the core and a different
$U(1)\ltimes\ZZ_2$ at infinity. However, the crucial
difference is, that our 'two' $U(1)$ gauge groups do not commute with
each other.

Interestingly, there is also another solution to the field equations,
which we briefly like to discuss. If $\chi=0$ there is a solution with
$m(r)=0$. After a rescaling of $q(r)$ one finds exactly the same
equations as were obtained in the Nielsen Olesen (NO) model by
\cite{devega} for the minimal flux $n=1$, provided we set the value of
$\lambda = 2 \xi$. Numerical solutions to these equations have been
studied before. For a special value of $\lambda$ one obtains the
solutions by solving, Bogomol'nyi type, first order equations,
signaling the possibility of extending the model to a super-symmetric
one whose super-symmetry increases for this value of the
parameter. The residual symmetry of this solution in our model is
$\ZZ_2$. One may wonder whether in our model this is a stable
solution. In the case of $\gamma =0$, i.e. $\chi=0$, the potential
(\ref{potential}) has the form:
\begin{equation} 
V=-\mu^2 X + \lambda X^2
\end{equation}
The minimum of this potential is obviously given by
$X=\frac{\mu^2}{2\lambda}$, with $X= \frac{1}{2}Tr(\Phi)^2$. Written
in the components $m$ and $q$ this gives:
\begin{equation} 
c_1 m^2 + c_2 q^2 =1
\end{equation}
with $c_1,c_2 >0$. A simple rescaling of $m$ and $q$ yields: $ m^2 +
q^2 =1$. As we require finite energy, this is one of the boundary
conditions for the fields $m$ and $q$ at spatial infinity. We now see 
that the boundary condition of the new solution, $m=0$, may be
continuously changed to one where $q=0$. But then the higgs field does
no longer stabilize the flux, which means that the flux will decay by
spreading out and losing more and more energy. Through this
process we end up in a ``rotated'' Alice phase of the theory where the
$U(1)$ generator points in the internal direction of the flux we
started of with. When going from the $m=0$ to the $q=0$ boundary
condition we pass an Alice phase whose $U(1)$ generator is
perpendicular to the internal direction of the flux we started
with. The upshot of these observations is that, if one wants to have a
stable Alice flux in an Alice phase, one needs to have $\gamma\neq
0$. Or, stated the other way around, if we do have $\gamma = 0$ and
are in an Alice phase with an (unstable) Alice flux, the Alice flux
will decay and change the Alice phase into another Alice phase, whose
$U(1)$ generator is in the internal direction of the flux we started
of with. This concludes what we have to say about conventional Alice
electrodynamics, in the remaining sections of the paper we will focus on
some alternative Alice models.

\section{Alternative Alice models.}
In this section we introduce two alternative models, which exhibit an
Alice electrodynamic phase. In these alternative models we choose the
higgs field(s) in the adjoint ($3$) representation of $SO(3)$. This
obviously means that the higgs field is not single valued in the
presence of an Alice flux, but this can be ``solved'' in two more
or less similar ways. One way is to put the internal space of the
higgs field ($X$) on a $\ZZ_2$ orbifold, i.e. $X$ and $-X$ are
identified with each other. The other way is to use (at least) two
higgs fields and put the total internal space of these two higgs
fields ($X$ and $Y$) on a $S_2$ orbifold, i.e. you identify the points
$(X,Y)$ and $(Y,X)$.\\ The action we use for both models is given by:
\begin{eqnarray}
S &=& \int d^4 x\ \{ Tr\ \{ -\frac{1}{4}F^{\mu\nu}F_{\mu\nu} -
\frac{1}{2}D^\mu X D_\mu X \ - \frac{1}{2}D^\mu Y D_\mu Y \ +
\frac{\hat{\gamma}}{2}[X,Y]^2 \}\nonumber\\&&+
\frac{\hat{\lambda}}{4}\{Tr\{X^2 + Y^2\}-f^2\}^2 \}
\label{action2}
\end{eqnarray}
Both theories allow the presence of an Alice flux. In the $S_2$ model
it means that one studies the twisted sector of the theory.

\subsection{Alice flux solutions}
\begin{figure}[t,h,b]
\psfig{figure=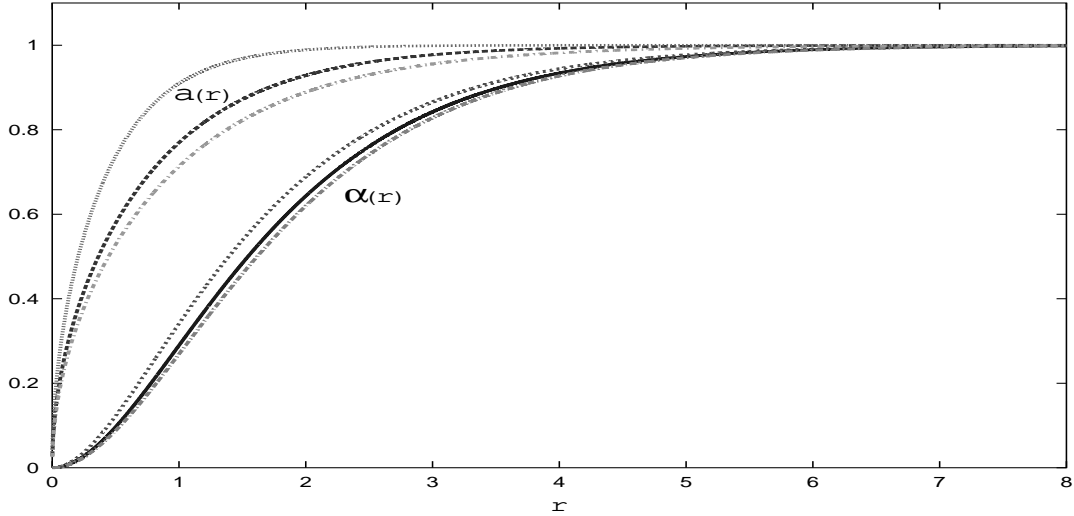,angle=-90,width=15cm,height=7cm}
\caption[somethingelse] {\footnotesize The fields $\alpha(r)$ and
$a(r)$ for $\lambda=2.0...0.5....0.3$.}
\label{jelper1.ps}
\end{figure}

\begin{figure}[t,h,b]
\psfig{figure=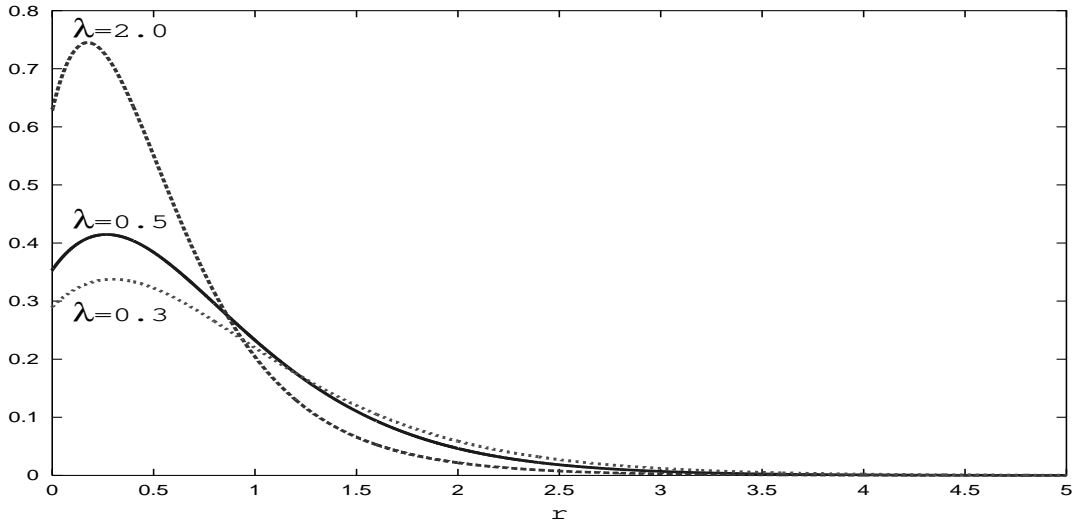,angle=-90,width=15cm,height=7cm}
\caption[somethingelse] {\footnotesize The energy density times $r$ of
the Alice flux for $\lambda=2.0...0.5....0.3$. }
\label{jelper2.ps}
\end{figure}

We now turn to the construction of regular cylindrically symmetric
(numerical) solutions corresponding to an Alice flux. At spatial infinity
one has $D_\theta X=0$, implying that the higgs field should have the
following form at spatial infinity.
\begin{equation}
X(\theta) = S(\theta) X(0) S^{-1}(\theta)
\end{equation}
with $S(\theta)$ the same as in (\ref{Stheta}). The flux associated with
$S(\theta)$, is topologically stable if the element of the gauge group
associated with $S(2\pi)$ is an element of the disconnected part of
the residual gauge group. A simple choice
\footnote{At this point we can not yet say that this is an element of the
disconnected part of the residual gauge group, but this will be done
consistently below.} is: $A_\theta=\frac{1}{2er}T_1$. This puts the
Alice flux in the internal $T_1$ direction. Writing the higgs field as
$X=x^a T_a$, it follows that $X(\theta)$ has the following form:\\
\begin{displaymath} 
X(\theta)= e^\frac{T_1\theta}{2}
\left(\begin{array}{ccc}
x_1(0)\\
x_2(0)\\
x_3(0)
\end{array}\right)
\end{displaymath}
Thus you see that $X(2\pi)$ is given by:
\begin{displaymath} 
X(2\pi)= 
\left(\begin{array}{ccc}
x_1(0)\\
-x_2(0)\\
-x_3(0)
\end{array}\right)
\end{displaymath}
The same holds, of course, for the other higgs field. Because the two
models differ slightly in constructing the ansatz, we will treat them
separately for the moment .\\
\\
\textbf{The $\ZZ_2$ model:}\\
For the $\ZZ_2$ model the boundary condition specified above implies that
either $x_1=0$ or $x_2=x_3=0$. Only in the first case, however, is
$S(2\pi)$ an element of the disconnected part of the gauge group. Thus
we have to put $x_1=0$. Later we will see that this choice is
important in order to obtain first order equations. At this point it is
convenient to introduce a different basis for the generators of the
gauge group, a basis naturally linked to the orientation of the
higgs field. Its elements are given by:\\
\begin{equation} 
S_a(\theta)=e^\frac{T_1\theta}{2} T_a e^\frac{-T_1\theta}{2}
\label{Sa}
\end{equation}
Now we write the higgs field as $X=x^a S_a$, where also in this
language one has to put $x_1=0$ to secure the possibility of a
topological stable solution.\\ In this model a single higgs field
would suffice, but for reasons of similarity we will use two. Our
ansatz than reads:
\begin{eqnarray}
A_\theta&=&\frac{\alpha(r)}{2er}S_1\\
X&=&a(r) S_3 \\
Y&=&c(r) S_1
\label{ansatz}
\end{eqnarray}
\\ \\ \textbf{The $S_2$ model:} \\ The 'double valuedness' is only
allowed if one uses an orbifold interpretation. So we impose a
strict relation between $X$ and $Y$.\\
\begin{displaymath}
X(\theta+2\pi)= 
\left(\begin{array}{ccc}
 x_1(\theta)\\
-x_2(\theta)\\
-x_3(\theta)
\end{array}\right) =
\left(\begin{array}{ccc}
 y_1(\theta)\\
 y_2(\theta)\\
 y_3(\theta)
\end{array}\right) =
Y(\theta)
\end{displaymath} 
Leading to: $x_1=y_1$ , $x_2=-y_2$ and $x_3=-y_3$. Again we are
going to work with the twisted generators (\ref{Sa}). A consistent
ansatz is the following one:
\begin{eqnarray}
A_\theta &=& \frac{\alpha(r)}{2er} S_1\\
X&=& a(r)S_3 + c(r)S_1\\
Y&=& -a(r)S_3 + c(r)S_1
\end{eqnarray}
For both cases one may insert the appropriate ansatz in the
field equations. This yields after a suitable rescaling, the same set
of differential equations for both models:
\begin{eqnarray}
\partial_r^2 \alpha(r)- \frac{1}{r}\partial_r \alpha(r) &=&
 (\alpha(r)-1)a^2(r) \label{eom11} \\ \partial_r^2 a(r) +
 \frac{1}{r}\partial_r a(r) &=& \frac{(\alpha(r)-1)^2}{4r^2}a(r) +
 \lambda a(r) (a^2(r)+c^2(r)-1)\nonumber\\ &&+\gamma c^2(r)a(r)
 \label{eom12} \\ \partial_r^2 c(r) + \frac{1}{r}\partial_r c(r)
 &=&\gamma a^2(r)c(r) + \lambda c(r) (a^2(r)+c^2(r)-1)
\label{eom13}
\end{eqnarray}
The asymptotic values of the fields are are as follows:\\[2mm]
\begin{displaymath}
\begin{tabular}{ccc}
\hline
field&$r\to0$ & r$ \to \infty$\\
\hline
$\alpha(r)$ & $0$ & $1$\\
$a(r)$ &$ 0$ &$1$\\
$c(r)$ &$constant$  & $0$ \\
\hline  
\end{tabular}
\end{displaymath}\\[2mm]
The boundary conditions are such that $S(2\pi)$ is an element of the
disconnected part of the residual gauge group.\\ We have constructed
numerical solutions to these equations, for different values of
$\lambda$ (and $\gamma$), with the use of a ``shooting'' method, see
figures \ref{jelper1.ps} and \ref{jelper2.ps}. As a matter of
fact, we only found solutions for which $c(r)=0$, although our
starting values were chosen quite general. This implies that there is
no dependence of the solutions we found, on $\gamma$.

In fact if $c(r)=0$, the equations become the same as in the case of a
NO flux with the critical value of the Landau coupling parameter
leading to first order Bogomol'nyi equations. However, there is an
important difference with the NO case. The ``winding'' number of the
Alice flux is fractional and equals $n= \frac{1}{2}$, a value which is
not admissible in the NO model. This is clearly a consequence of the
different breaking schemes of the theories in question.\\ There is a
special role in these theories for the parameter $\gamma$, if we set
$\gamma=0$ the equations are very similar to the equations
(\ref{eom21})-(\ref{eom23}) with $\chi=0$ . Though $\gamma$ appears to
play no role as long as $c(r)=0$, this is not quite the case. We don't
want  $\gamma$ to vanish because than we run more or less into the same
problem as in the conventional model for Alice electrodynamics with
$\chi=0$. The solution with $c(r)=0$ would still be a solution of the
field equations, but the flux would no longer be stable. It would
be allowed to decay into the vacuum. In fact in the alternative models it
is quite clear what happens at $\gamma=0$, the potential term
proportional to $\gamma$ -- assuming it is nonzero -- ensures that there
is no continuous path in the vacuum manifold connecting the $c=0$ to
the $a=0$ boundary condition\footnote{Remember, there is also the boundary condition $a^2+c^2=1$.}. If $\gamma=0$ such a path does exist.

There is a simple relation between the $\ZZ_2$ model with one higgs
field and the $S_2$ model with two higgs fields, in the presence of an
Alice string. In the presence of an Alice string the field component
of the higgs field parallel to the Alice flux is zero in the $\ZZ_2$
model, whereas in the $S_2$ model this is in general only true far
away from the core. So, in some sense the $\ZZ_2$ model is a long
wavelength approximation of the $S_2$ model, but remarkably enough,
it does support solutions which are regular everywhere nevertheless. The
action of both models becomes the same, up to a rescaling, if the
components parallel to the Alice flux, of the higgs fields in the
$S_2$ model, are set equal to zero.


\subsection{First order equations}
As mentioned before, if one sets $c(r)=0$, the set of equations,
(\ref{eom11}-\ref{eom13}), reduces to the same set that one would obtain in the
NO model for a solution with winding number $n=\frac{1}{2}$ . It thus
appears that one can, in the sector that contains a topologically
stable Alice flux, project both theories on a sector of the NO
model. This raises the question whether it would be possible to find
first order equations in both models. In the $\ZZ_2$ model, with
only a single higgs field, this projection is the clearest. For the
rest of this section we will therefore concentrate on this case.

One of the features of the NO theory is that for a certain value of
the coupling constant $\lambda$, the solutions can be obtained from
first order equations. These first order equations can be found {\em \`a
la} Bogomol'nyi, by rewriting the energy density as a sum of squares
plus a topological term. In the case of static solutions, the energy
density and the Lagrangian differ only by a sign, implying that extrema
of the energy are also extrema of the Lagrangian. Consequently,
solutions of minimal energy are stable static solutions of the full
set of second order field equations. The energy
of the $\ZZ_2$ model is given by:
\begin{equation}
E=\frac{1}{2} \int d^3 x \{~ Tr (~ E_i^2 + B_i^2 + (D_i X)^2 + (D_t
X)^2 ~) + \frac{\lambda}{2}(~ (x^1)^2 + (x^2)^2 + (x_3)^2-1)~ \}
\end{equation}
In the static case with no electric fields, in the gauge $A_t=0$, one
has $E_i = \partial_t A_i = 0$ and $D_t X = \partial_t X = 0$,
reducing the expression to, 
\begin{equation}
E=\frac{1}{2} \int d^3 x \{~ Tr (~ B_i^2 + (D_i X)^2 ~) +
\frac{\lambda}{2}( (x^1)^2 + (x^2)^2 + (x_3)^2-1) ~\}
\end{equation}
Restricting ourselves to the plane, the energy written in components
is given by:
\begin{eqnarray} 
E&=&\frac{1}{2}\int d^2x\{ (B_z^1)^2 + (B_z^2)^2 + (B_z^3)^2 + ((D_\nu
X)^1)^2 + ((D_\nu X)^2)^2 + ((D_\nu X)^3)^2 \nonumber\\&& +
\frac{\lambda}{2} ( (x^1)^2 + (x^2)^2 + (x^3)^2-1)\},
\label{energy}
\end{eqnarray}
where now the upper label refers to the internal directions (in the
normal, non-twisted basis) and the lower label to the spatial
directions. From this expression we are unable to obtain first order
equations, however, if we restrict ourselves to the subspace of
solutions containing an Alice flux, there is something we can
do. Let's call the internal direction in which the Alice flux 'points'
the '$1$' direction, so if one is looking for topological stable fluxes
one needs to have $x_1=0$, as argued before. In that case we may
write the energy as,
\begin{eqnarray} 
E&=&\frac{1}{2}\int d^2x\{ (B_z^1)^2 + (B_z^2)^2 + (B_z^3)^2 +
([A_1,X]^1)^2 + ([A_2,X]^1)^2 + (\partial_1 x^2 +A^1_1 x^3)^2
\nonumber\\&&+ (\partial_1 x^3 - A^1_1 x^2)^2+ (\partial_2 x^2 + A^1_2
x^3)^2 + (\partial_2 x^3 +A^1_2 x^2)^2\nonumber\\&& +
\frac{\lambda}{2}( (x^2)^2 + (x^3)^2 - 1)\}
\end{eqnarray}
For the case of $\lambda=\frac{1}{2}$ this can be brought into the
form:
\begin{eqnarray}
E&=&\frac{1}{2}\int d^2x\{ (B_z^2)^2 + (B_z^3)^2 + ([A_1,X]^1)^2 +
([A_2,X]^1)^2 \nonumber\\&& + (B_z^1\pm\frac{1}{2}( (x^2)^2 + (x^3)^2
-1) )^2 + ((\partial_1x^2+A_1^1x^3)\mp(\partial_2x^3-A_2^1x^2))^2
\nonumber\\&&+ ((\partial_2x^2+A_2^1x^3)\pm(\partial_1x^3-A_1^1x^2))^2
\pm [A_1,A_2]^1( (x^2)^2 + (x^3)^2 - 1) \}\nonumber\\&& \pm
\frac{1}{2} \int d^2x (B_z + [A_1,A_2])^1 
\end{eqnarray}
Still, there appears to be a problem, because there are two terms in
this expression of the energy density, which are not squares and as we
will show later, only one of them is proportional to the winding
number. The other term, $[A_1,A_2]^1( (x^2)^2 + (x^3)^2 - 1)$,
therefore appears to be problematic. This problem can fortunately be
cured rather straightforwardly. Remember that we are already in a 'gauge'
$A_t=0$. where furthermore the static fields are time independent. In
this situation the residual gauge freedom of time independent gauge
transformations may be used to put the term $[A_1,A_2]^1$ equal to
zero. In this gauge the energy density consists only of squares and a term
proportional to the winding number. The minimum of the energy is
now easily obtained by putting all squares in the energy density equal
to zero. This then yields a set of first order equations of which
the solutions are also solutions of the full field equations. The
first order equations, including the gauge conditions, are:
\begin{eqnarray}
[A_1,A_2]^1&=&0\\
B_z^2 &=& 0\\
B_z^3 &=& 0\\
{[A_1,X]}^1 &=& 0\\
{[A_2,X]}^1 &=& 0\\
(B_z^1\pm\frac{1}{2}( (x^2)^2 + (x^3)^2 -1)&=&0\\
(\partial_1x^2+A_1^1x^3)\mp(\partial_2x^3-A_2^1x^2)&=&0\\
(\partial_2x^2+A_2^1x^3)\pm(\partial_1x^3-A_1^1x^2)&=&0
\end{eqnarray}
The last three equations are identical to those that were obtained in
the NO model. The energy of solutions to this set of equations are
fully determined by the term $\int d^2x (B_z + [A_1,A_2])^1$ which is
proportional to the winding number, as we show next.

The general expression for $X$ in the presence of an Alice flux in the
first isospin direction ``along'' $T_1$, becomes:
\begin{equation}
X=x(r)e^{2\pi i \chi(\theta) T_1} T_2 = x(r)\cos(2\pi\chi(\theta))T_2
+ x(r)\sin(2\pi\chi(\theta))T_3 \equiv aT_2 + bT_3
\end{equation}
With $\chi(\theta + 2\pi) = \chi(\theta) +\frac{1}{2}$. For $r\to
\infty$ one has $x(r\to\infty)=1$, and the winding number can be extracted
from the asymptotics by:
\begin{equation}
\frac{-i}{2\pi} \oint^{r\to\infty} d\ln(a+i*b) = n = \frac{1}{2}
\end{equation}
For $r\to \infty$ one also has the spatial covariant derivatives $DX=0$ or:
\begin{equation}
\partial X= [A,X]
\end{equation}
or in components:
\begin{displaymath}
\begin{array}{cl}
[A,X]^1  &= 0\\
\partial a &= -A^1 b\\
\partial b &= A^1 a
\end{array}
\end{displaymath}
From this one finds $\partial\ln(a+i*b) = i*A^1$, which means that:
\begin{equation}
\frac{-i}{2\pi} \oint^{r\to\infty} d\ln(a+i*b) =
\frac{1}{2\pi}\oint^{r\to\infty} d\vec{l}~ \hat{l}\cdot A^1 =
\frac{1}{2\pi}\int d^2x (B_z^1 + [A_1,A_2]^1)
\end{equation}
Thus the rescaled energy of the solutions is equal to
$\frac{\pi}{2}$. Note that the above expressions do not look gauge invariant
because we are evaluating a gauge invariant expression in a particular
gauge.\\ One should, of course, check whether the first order
equations actually do have any solutions of the type we are interested
in. By inserting the ansatz used before and putting $c(r)$ to zero,
one arrives at the following set of coupled non-linear first order
equations.
\begin{eqnarray}
r\partial_ra(r)&=&\frac{1}{2}(1-\alpha(r))a(r)\\
\frac{1}{r}\partial_r \alpha(r)&=& 1-a^2(r)
\end{eqnarray}
Now, these turn out to be a special case of the equations encountered before by
De Vega and Schaposnik \cite{devega} in their study of the NO
model. They where obviously only interested in the case of integer
winding number, whereas we are interested in the case of fractional
winding number $n=\frac{1}{2}$. The corresponding numerical solution
is given in figures \ref{jelper1.ps} and \ref{jelper2.ps}.

We have attained our goal of obtaining a set of first order equations,
of which the solutions are also static minima of the energy (with no
electric fields). As is well-known , first order equations play a 
deep role in gauge theories. Bogomol'nyi \cite{bogo} explained,
for the NO model, that solutions which come from the first order
equations are also minima of the energy, which implied the neutral
stability of such solutions. Later it was shown \cite{hlousek} that
the occurrence so-called Bogomol'nyi equations is tightly connected to
the existence of a super-symmetric extension of the theory. The
explicit super symmetry extension of the NO model was given by \cite{edelstein}
and in agreement with \cite{hlousek} showed that the first order
equations indeed follow from an increase of super-symmetry. In our
models we also found first order equations whose solutions are also
solutions to the full set of second order field equations. We showed
that the solutions are also minima of the energy. This obviously
raises the question if these first order solutions can also be
explained by an increase of super symmetry of a super-symmetric extension
of our models. A superficial analysis suggests that this is not the
case, basically because we can only recover the Bogomol'nyi argument
within the context of a very restrictive ansatz. In this respect the
situation is similar to that encountered in the study of regular $\ZZ
_N $ monopoles \cite{bais}.

\section{Conclusion}
In this paper we proposed two new models which both possess an Alice
electrodynamics phase. For both models we constructed solutions
corresponding to a topologically stable Alice flux. We found a way to
project the theories on the Nielsen Olesen model and, in that way,
obtained first order equations. Solutions to these first order
equations corresponding to minima of the energy (without electric
fields) were constructed numerically.\\ We close with a brief remark
concerning the zero modes of our solution. E.Weinberg \cite{weinberg}
showed that in the NO model, for the critical value
$\lambda=\frac{1}{2}$, a flux with winding number $n$ has $2n$ zero
modes. These modes are interpreted as being the positions of the unit
fluxes. At first sight this appears to give problems for the case of
$n=\frac{1}{2}$, but carefully redoing section IV of the article
mentioned, in particular using the fact that our fields are allowed to
be double valued, one may show that the answer for $n=\frac{1}{2}$ is
that there are again two zero modes, as one would expect.\\ We thank
M.M.H.Postma \cite{postma} for his contributions in the early stages
of the project.


\begin{thebibliography}{99}

\bibitem{NO} H.B.Nielsen, P.Olesen, Nucl. Phys. B 61 (1973) 45.
\bibitem{schwarz} A.S.Schwarz, Nucl. Phys. B 208 (1982) 141.
\bibitem{Alford} M.Alford, K.Benson, S.Coleman, J. March-Russell, F.Wilczek, Nucl. Phys. B 349 (1991) 414. 
\bibitem{shankar} R.Shankar, Phys. Rev. D 14 (1976) 1107.
\bibitem{postma} M.M.H.Postma, Master's thesis, University of Amsterdam (1997).
\bibitem{witten} E.Witten, Nucl. Phys. B 249 (1985) 557.
\bibitem{devega} H.J.de Vega, F.A.Schaposnik, Phys. Rev. D 14 (1976) 1100.
\bibitem{bogo}  Bogomol'nyi, Sov.J.Nucl.Phys 24 (1976) 449. 
\bibitem{hlousek} Z.Hlousek, D.Spector, Nucl. Phys. B 397 (1993) 173.
\bibitem{edelstein} J.Edelstein, C.Nunez, F.Schaposnik, Phys. Lett. B 329 (1994) 39
\bibitem{bais} F.A.Bais, R.Laterveer, Nucl. Phys. B 307 (1988) 487. 
\bibitem{weinberg} E.J.Weinberg, Phys. Rev. D 19 (1979) 3008
\end{thebibliography}
\end{document}